\documentclass{webofc}
\usepackage[varg]{txfonts}   
\begin{document}

\title{Exploring the millimetre emission in nearby galaxies:}
%
%
\subtitle{analysis of the edge-on galaxy NGC~891}

\author{\firstname{S.}~\lastname{Katsioli}\inst{\ref{NOA}, \ref{UOA}}\fnsep\thanks{\email{s.katsioli@noa.gr}}
  \and \firstname{R.}~\lastname{Adam} \inst{\ref{LLR}}
  \and  \firstname{P.}~\lastname{Ade} \inst{\ref{Cardiff}}
  \and  \firstname{H.}~\lastname{Ajeddig} \inst{\ref{CEA}}
  \and  \firstname{P.}~\lastname{Andr\'e} \inst{\ref{CEA}}
  \and \firstname{E.}~\lastname{Artis} \inst{\ref{LPSC}}
  \and  \firstname{H.}~\lastname{Aussel} \inst{\ref{CEA}}
  \and  \firstname{A.}~\lastname{Beelen} \inst{\ref{IAS}}
  \and  \firstname{A.}~\lastname{Beno\^it} \inst{\ref{Neel}}
  \and  \firstname{S.}~\lastname{Berta} \inst{\ref{IRAMF}}
  \and  \firstname{L.}~\lastname{Bing} \inst{\ref{LAM}}
  \and  \firstname{O.}~\lastname{Bourrion} \inst{\ref{LPSC}}
  \and  \firstname{M.}~\lastname{Calvo} \inst{\ref{Neel}}
  \and  \firstname{A.}~\lastname{Catalano} \inst{\ref{LPSC}}
  \and  \firstname{I.}~\lastname{De~Looze} \inst{\ref{Ghent}, \ref{UCL}}
  \and  \firstname{M.}~\lastname{De~Petris} \inst{\ref{Roma}}
  \and  \firstname{F.-X.}~\lastname{D\'esert} \inst{\ref{IPAG}}
  \and  \firstname{S.}~\lastname{Doyle} \inst{\ref{Cardiff}}
  \and  \firstname{E.~F.~C.}~\lastname{Driessen} \inst{\ref{IRAMF}}
  \and  \firstname{G.}~\lastname{Ejlali} \inst{\ref{Iran}}
  \and  \firstname{M.}~\lastname{Galametz} \inst{\ref{CEA}}
  \and  \firstname{F.}~\lastname{Galliano} \inst{\ref{CEA}}
  \and  \firstname{A.}~\lastname{Gomez} \inst{\ref{CAB}}
  \and  \firstname{J.}~\lastname{Goupy} \inst{\ref{Neel}}
  \and  \firstname{A.~P.}~\lastname{Jones} \inst{\ref{IAS}}
  \and  \firstname{A.}~\lastname{Hughes} \inst{\ref{IAS}}
  \and  \firstname{F.}~\lastname{K\'eruzor\'e} \inst{\ref{LPSC}}
  \and  \firstname{C.}~\lastname{Kramer} \inst{\ref{IRAMF},\ref{IRAME}}
  \and  \firstname{B.}~\lastname{Ladjelate} \inst{\ref{IRAME}}
  \and  \firstname{G.}~\lastname{Lagache} \inst{\ref{LAM}}
  \and  \firstname{S.}~\lastname{Leclercq} \inst{\ref{IRAMF}}
  \and  \firstname{J.-F.}~\lastname{Lestrade} \inst{\ref{LERMA}}
  \and  \firstname{J.-F.}~\lastname{Mac\'ias-P\'erez} \inst{\ref{LPSC}}
  \and  \firstname{S.~C.}~\lastname{Madden} \inst{\ref{CEA}}
  \and  \firstname{A.}~\lastname{Maury} \inst{\ref{CEA}}
  \and  \firstname{P.}~\lastname{Mauskopf} \inst{\ref{Cardiff},\ref{Arizona}}
  \and \firstname{F.}~\lastname{Mayet} \inst{\ref{LPSC}}
  \and  \firstname{A.}~\lastname{Monfardini} \inst{\ref{Neel}}
  \and  \firstname{M.}~\lastname{Mu\~noz-Echeverr\'ia} \inst{\ref{LPSC}}
  \and  \firstname{A.}~\lastname{Nersesian} \inst{\ref{Ghent}, \ref{NOA}}
  \and  \firstname{L.}~\lastname{Perotto} \inst{\ref{LPSC}}
  \and  \firstname{G.}~\lastname{Pisano} \inst{\ref{Cardiff}}
  \and  \firstname{N.}~\lastname{Ponthieu} \inst{\ref{IPAG}}
  \and  \firstname{V.}~\lastname{Rev\'eret} \inst{\ref{CEA}}
  \and  \firstname{A.~J.}~\lastname{Rigby} \inst{\ref{Cardiff}}
  \and  \firstname{A.}~\lastname{Ritacco} \inst{\ref{IAS}, \ref{ENS}}
  \and  \firstname{C.}~\lastname{Romero} \inst{\ref{Pennsylvanie}}
  \and  \firstname{H.}~\lastname{Roussel} \inst{\ref{IAP}}
  \and  \firstname{F.}~\lastname{Ruppin} \inst{\ref{MIT}}
  \and  \firstname{K.}~\lastname{Schuster} \inst{\ref{IRAMF}}
  \and  \firstname{S.}~\lastname{Shu} \inst{\ref{Caltech}}
  \and  \firstname{A.}~\lastname{Sievers} \inst{\ref{IRAME}}
  \and  \firstname{M.~W.~L.}~\lastname{Smith} \inst{\ref{Cardiff}}
  \and  \firstname{F.}~\lastname{Tabatabaei} \inst{\ref{Iran}, \ref{Laguna}}
  \and  \firstname{C.}~\lastname{Tucker} \inst{\ref{Cardiff}}
  \and  \firstname{E.~M.}~\lastname{Xilouris} \inst{\ref{NOA}}
  \and  \firstname{R.}~\lastname{Zylka} \inst{\ref{IRAMF}}
}
  
  \institute{
    National Observatory of Athens, Institute for Astronomy, Astrophysics, Space Applications and Remote Sensing, Ioannou Metaxa and Vasileos Pavlou GR-15236, Athens, Greece
    \label{NOA}
    \and
    Department of Astrophysics, Astronomy \& Mechanics, Faculty of Physics, University of Athens, Panepistimiopolis, GR-15784 Zografos, Athens, Greece
    \label{UOA}
    \and
    LLR (Laboratoire Leprince-Ringuet), CNRS, École Polytechnique, Institut Polytechnique de Paris, Palaiseau, France
    \label{LLR}
    \and
    School of Physics and Astronomy, Cardiff University, Queen’s Buildings, The Parade, Cardiff, CF24 3AA, UK 
    \label{Cardiff}
    \and
    AIM, CEA, CNRS, Universit\'e Paris-Saclay, Universit\'e Paris Diderot, Sorbonne Paris Cit\'e, 91191 Gif-sur-Yvette, France
    \label{CEA}
    \and
    Univ. Grenoble Alpes, CNRS, Grenoble INP, LPSC-IN2P3, 53, avenue des Martyrs, 38000 Grenoble, France
    \label{LPSC}
    \and
    Universit\'e Paris-Saclay, CNRS, Institut d'astrophysique spatiale, 91405, Orsay, France
    \label{IAS}
    \and
    Institut N\'eel, CNRS, Universit\'e Grenoble Alpes, France
    \label{Neel}
    \and
    Institut de RadioAstronomie Millim\'etrique (IRAM), Grenoble, France
    \label{IRAMF}
    \and
    Aix Marseille Univ, CNRS, CNES, LAM (Laboratoire d'Astrophysique de Marseille), Marseille, France
    \label{LAM}
    \and
    Sterrenkundig Observatorium Universiteit Gent, Krijgslaan 281 S9, B-9000 Gent, Belgium
    \label{Ghent}
    \and
    Department of Physics and Astronomy, University College London, Gower Street, London WC1E 6BT, UK
    \label{UCL}
    \and 
    Dipartimento di Fisica, Sapienza Universit\`a di Roma, Piazzale Aldo Moro 5, I-00185 Roma, Italy
    \label{Roma}
    \and
    Univ. Grenoble Alpes, CNRS, IPAG, 38000 Grenoble, France 
    \label{IPAG}
    \and
    Institute for Research in Fundamental Sciences-IPM, Larak Garden,19395-5531 Tehran, Iran
    \label{Iran}
    \and
    Centro de Astrobiolog\'ia (CSIC-INTA), Torrej\'on de Ardoz, 28850 Madrid, Spain
    \label{CAB}
    \and  
    Instituto de Radioastronom\'ia Milim\'etrica (IRAM), Granada, Spain
    \label{IRAME}
    \and 
    LERMA, Observatoire de Paris, PSL Research University, CNRS, Sorbonne Universit\'e, UPMC, 75014 Paris, France 
    \label{LERMA}
    \and
    School of Earth and Space Exploration and Department of Physics, Arizona State University, Tempe, AZ 85287, USA
    \label{Arizona}
    Laboratoire de Physique de l’\'Ecole Normale Sup\'erieure, ENS, PSL Research University, CNRS, Sorbonne Universit\'e, Universit\'e de Paris, 75005 Paris, France 
    \label{ENS}
    \and
    Department of Physics and Astronomy, University of Pennsylvania, 209 South 33rd Street, Philadelphia, PA, 19104, USA
    \label{Pennsylvanie}
    \and 
    Institut d'Astrophysique de Paris, Sorbonne Université, CNRS (UMR7095), 75014 Paris, France
    \label{IAP}
    \and
    Kavli Institute for Astrophysics and Space Research, Massachusetts Institute of Technology, Cambridge, MA 02139, USA
    \label{MIT}
    \and
    Caltech, Pasadena, CA 91125, USA
    \label{Caltech}
    \and
    Instituto  de  Astrofísica  de  Canarias,  Vía  L’actea  S/N,  38205  La Laguna, Spain
    \label{Laguna}
  }

  \abstract{
New observations of the edge-on galaxy NGC~891, at 1.15 and 2~mm obtained with the IRAM~30-m telescope and the NIKA2 camera, within the framework of the IMEGIN (Interpreting the Millimetre Emission of Galaxies with IRAM and NIKA2) Large Program, are presented in this work. By using multiwavelength maps (from the mid-IR to the cm wavelengths) we perform SED fitting in order to extract the physical properties of the galaxy on both global and local ($\sim$kpc) scales. For the interpretation of the observations we make use of a state-of-the-art SED fitting code, HerBIE (HiERarchical Bayesian Inference for dust Emission). The observations indicate a galaxy morphology, at mm wavelengths, similar to that of the cold dust emission traced by sub-mm observations and to that of the molecular gas. The contribution of the radio emission at the NIKA2 bands is very small (negligible at 1.15~mm and $\sim10\%$ at 2~mm) while it dominates the total energy budget at longer wavelengths (beyond 5~mm). On local scales, the distribution of the free-free emission resembles that of the dust thermal emission while the distribution of the synchrotron emission shows a deficiency along the major axis of the disc of the galaxy.    


}
\maketitle
\section{Introduction}
\label{intro} 

Nearby galaxies have provided a great piece of knowledge in the field of galaxy formation and evolution. Their proximity allows us to study galaxies at high resolution and to derive their properties on local (sub-kpc/kpc) scales. The current study is part of the IMEGIN Large Program (P.I.: {\it S. Madden}), a NIKA2 guaranteed time program of 200~hours allocated telescope time with the aim of mapping 22 nearby galaxies (distances of~<~30~Mpc) at 1.15 and 2~mm using the NIKA2 continuum camera at the IRAM~30-m telescope \citep{NIKA2-performance, NIKA2-general, NIKA2-instrument, NIKA2-electronics}. The main objective of the project is to explore the emission at millimetre wavelengths originating from nearby galaxies, a complex wavelength regime with contributions from a mix of different emission mechanisms (free-free, synchrotron, and dust thermal emission). 
This is the first time that nearby galaxies are being observed in the continuum at mm wavelengths at resolutions of 11.1$^{\prime\prime}$ and 17.6$^{\prime\prime}$ ($\sim0.5$~kpc, and $\sim0.8$~kpc at a distance of 10~Mpc) at 1.15 and 2~mm respectively. At these scales it is possible to disentangle the contribution of the different emission mechanisms in different environments inside the galaxies (disc, halo, HII regions, etc). 

As a pilot study, we analyze the NIKA2 observations of the edge-on galaxy NGC~891, a SA(s)b type galaxy \cite{2013AJ....145..137K}. Due to its proximity (D~=~9.6~Mpc; \cite{2004ApJ...606..829S}) and its edge-on orientation (i~$\cong$~89.8$^\circ$; \cite{1999A&A...344..868X}), NGC~891 is one of the most observed and well studied galaxies of the Local Universe (e.g. \cite{1999A&A...344..868X, hughes, mulcahy, yoon}).

\section{Observations and Analysis}
\label{sec:obs}

\begin{figure}
\centering
\includegraphics[width=12cm,clip]{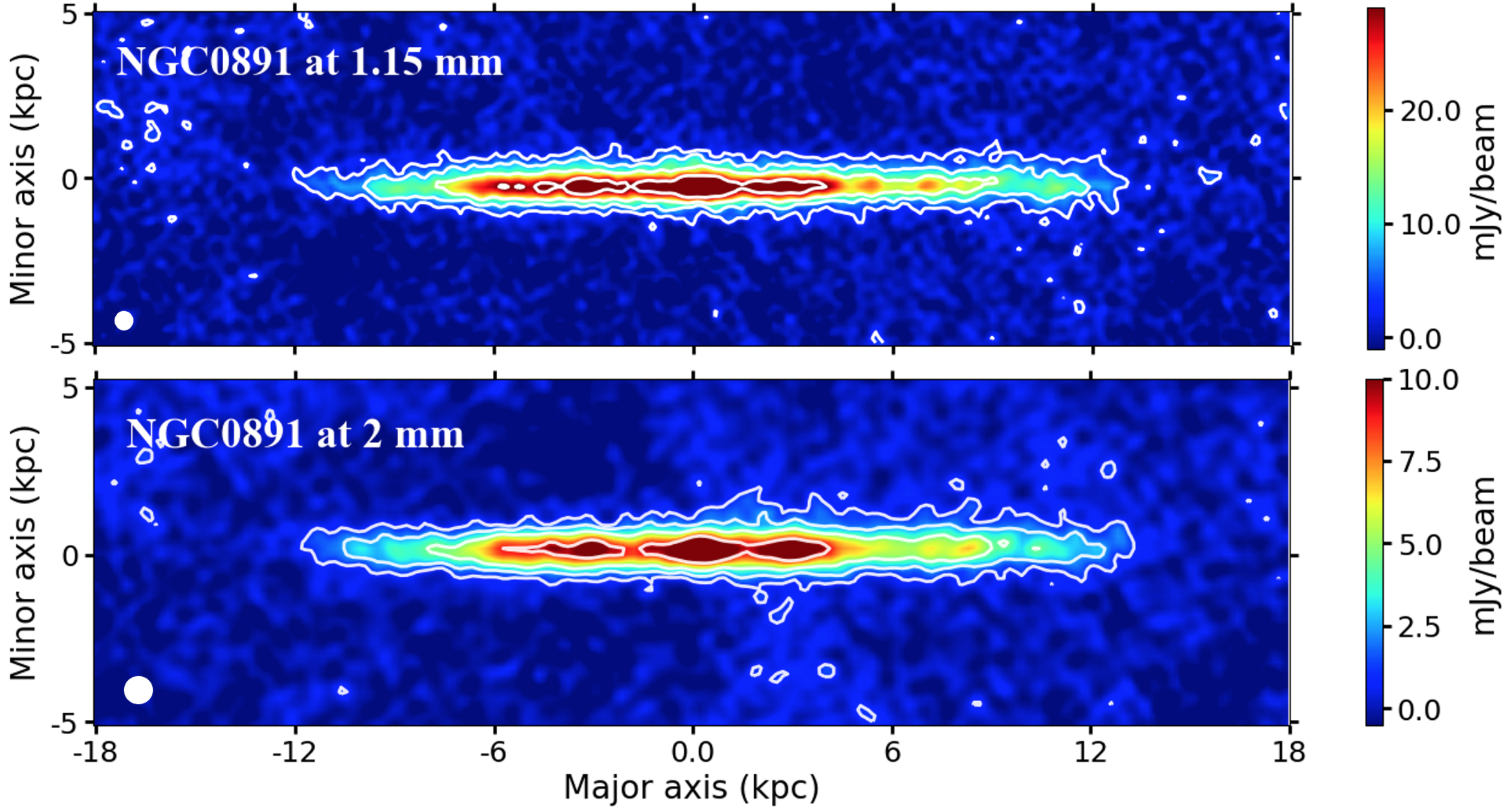}
\caption{NIKA2 high-resolution millimeter observations of NGC~891 at 1.15 (top panel) and 2~mm (bottom panel). The maps are presented in their native resolution (11.1$^{\prime\prime}$ and 17.6$^{\prime\prime}$ at 1.15 and 2~mm respectively; indicated by the white circles in each panel) and they are rotated by 22.9$^\circ$ (the position angle of the galaxy). Their sky RMS values are 1.0~mJy~beam$^{-1}$ and 0.3~mJy~beam$^{-1}$ at 1.15~mm and 2~mm. The surface brightness contours correspond to 3.5, 8, 15 and 30~$\times$~RMS.}
\label{fig1}       
\end{figure}
 
The NIKA2 observations of NGC~891 were obtained between October 2019 and January 2020 with a total on-source integration time of 7~hours. The maps of the galaxy at 1.15 and 2~mm were reduced using the  GILDAS/PIIC\footnote{https://publicwiki.iram.es/PIIC/} (version of 29.04.2020) software \citep{2013ascl.soft03011Z} and are shown in Fig.~\ref{fig1} (top and bottom panels respectively). The edge-on orientation of the galaxy and the high resolution of the maps allow us to discern different regions along the major axis with the main characteristics being a peak emission at the nucleus and secondary maxima at $\sim 3$~kpc either side of the center. These emission features closely resemble the distribution of the cold dust traced by sub-mm observations \cite{2011A&A...531L..11B} as well as of the molecular gas \citep{1993ApJ...404L..59S,1992A&A...266...21G} which are possibly originating either from a ring structure or limb brightening associated with spiral arms (see \cite{1998ApJ...507L.125A}).

In our study we performed a Spectral Energy Distribution (SED) analysis of NGC~891 using multiwavelength data ranging from 3.4~$\mu$m to 5~cm. We used archival data from the {\it WISE} telescope (3.4, 4.6, 11.6 and 22.1~$\mu$m), the {\it Spitzer} telescope (3.6, 4.5, 5.8, and 24~$\mu$m), as well as the {\it Herschel} telescope (70, 100, 160, 250, and 350~$\mu$m).
Along with the NIKA2 observations at 1.15 and 2~mm we also used centimeter data from the {\it AMI} (2~cm) and {\it EVLA} (5~cm) radio telescopes. The archival data have been retrieved from the DustPedia\footnote{http://dustpedia.astro.noa.gr/} and NED\footnote{https://ned.ipac.caltech.edu/} databases. The maps were all convolved (using Gaussian Kernels) to the same angular resolution of 25$^{\prime\prime}$ corresponding to the resolution of the 350~$\mu$m SPIRE map. This resolution corresponds to $\sim 1.1$~kpc at the distance of NGC~891. All maps have also been regridded into a common reference frame with a pixel size of 8$^{\prime\prime}$. Fitting the continuum emission of the galaxy, we had to remove the background sources from the near-IR maps as well as to correct the 1.15~mm map (see \cite{NIKA2-performance} for the NIKA2 transmission curves) from the CO(2-1) line contamination following the procedure described in \cite{CO_line}.
In order to compute the global fluxes at each wavelength we used an elliptical aperture centered at RA$_{J2000} = 2^h22^m33^s$, DEC$_{J2000}=+42^{\circ}20^{\prime}53^{\prime\prime}$ \cite{2007AJ....134.1019O} with major and minor axes of 5$^{\prime}$ and 48$^{\prime\prime}$ respectively and positional angle of 22.9$^\circ$, so as to 
encompass the bulk of the emission originating from the disc of the galaxy. Additional global photometry at radio wavelengths (from {\it Planck, 100-m Effelsberg, OVRO, WSRT} and {\it GBT} telescopes) were also culled from the literature.

\begin{figure*}[t]
\centering
\includegraphics[width=6.5cm, clip]{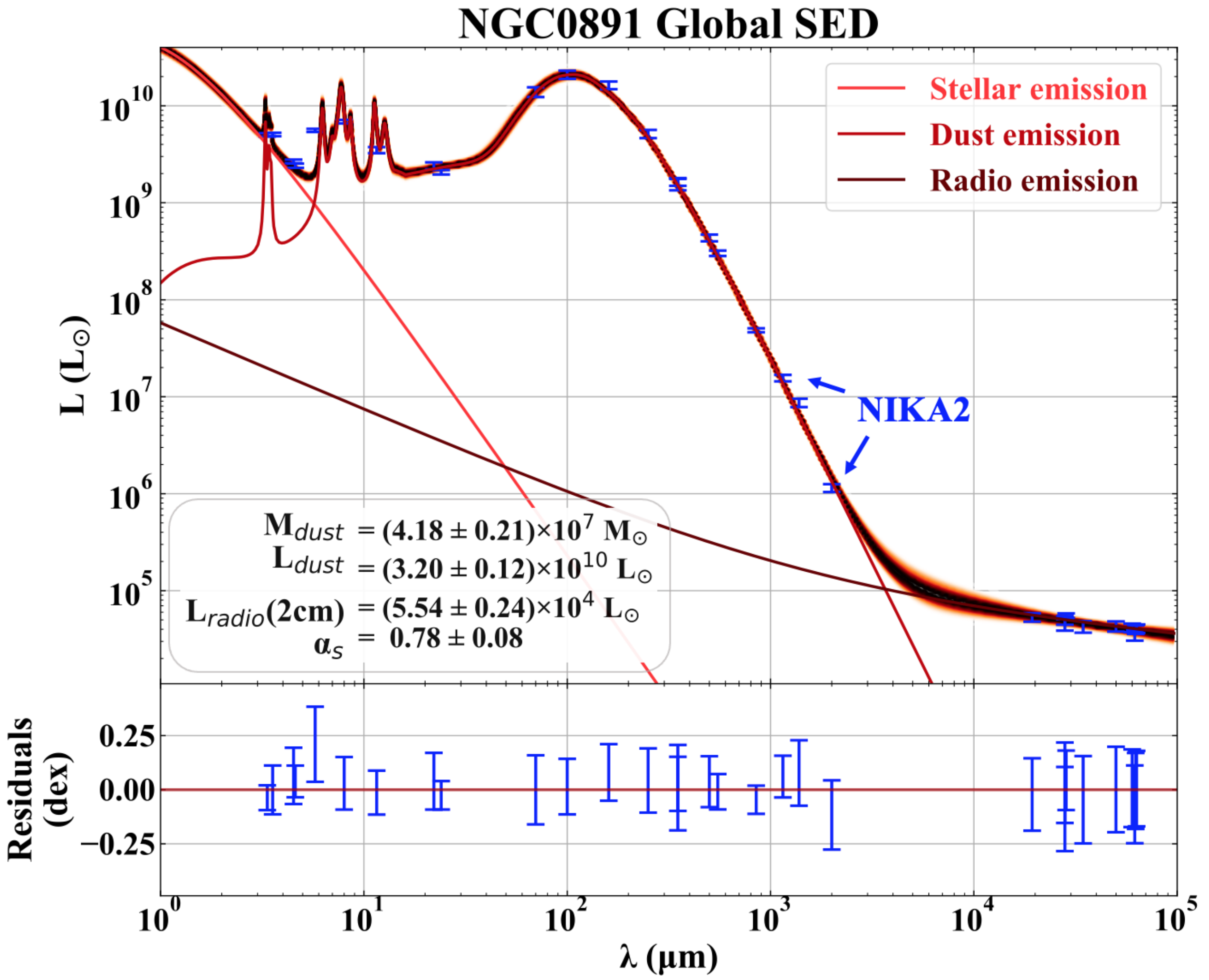}
\includegraphics[width=5.6cm, clip]{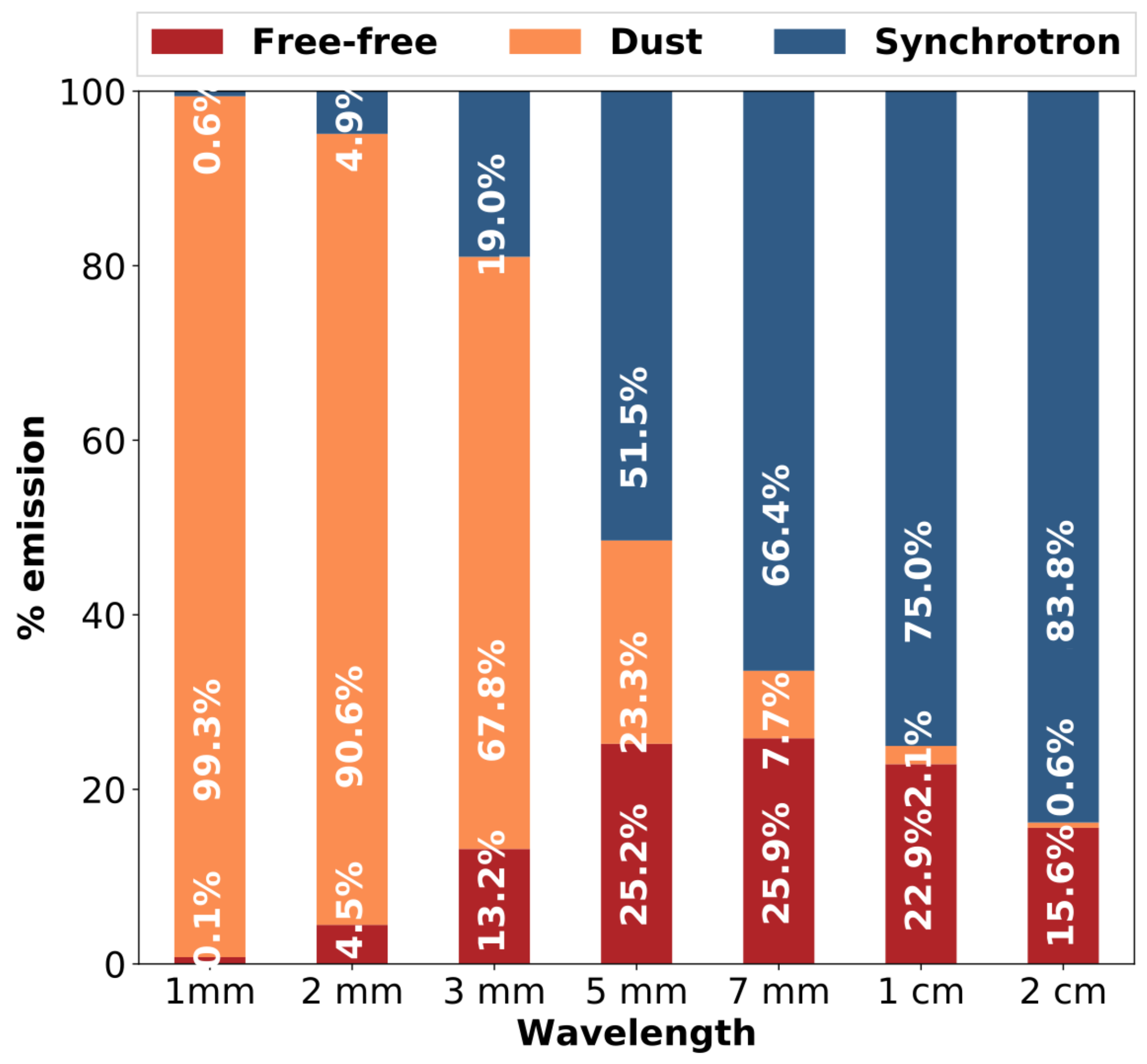}
\caption{Left panel: the global SED of NGC~891 with multiwavelength data ranging from 3.4~$\mu$m to 6.3~cm. The curve passing through the photometric data is the best fitted model with the width of the curve indicating the uncertainty of the model along the spectrum. The emission components of the total SED of the galaxy, stellar, dust and radio, are plotted with orange, red and brown color respectively. The residuals between the observed and the model predicted fluxes are given in the graph below the SED in dex. Right panel: An emission decomposition plot for characteristic wavelengths in the spectral range between 1~mm to 2~cm. The percentage of the dust emission is presented with orange color, while red and blue colors are used for percentages of the free-free and the synchrotron emission respectively.} 
\label{fig2}
\end{figure*}

To interpret the SED we used the SED fitting code HerBIE (Hierarchical Bayesian Inference for dust Emission; \cite{2018MNRAS.476.1445G, 2021A&A...649A..18G}). In contrast to the commonly used least-squares fitting method, HerBIE eliminates the noise-induced correlations between the parameters. The code takes into account realistic optical properties and stochastic heating for the dust grains by making use of the THEMIS dust model which is described in detail in \cite{2013A&A...555A..39J, 2017A&A...602A..46J}. Free-free and synchrotron emission mechanisms are also implemented in the code. The global SED of the galaxy, fitted by the HerBIE code, is presented in the left panel in Fig.~\ref{fig2}.  

With an angular resolution of 25$^{\prime\prime}$ (see Sec.~\ref{sec:obs}) the observations used in this study allowed us to perform spatially resolved SED fitting within the galaxy and create maps of the local physical properties associated with the dust and radio emission. 
The maps presented in Fig.~\ref{fig3} show the total dust luminosity map (left), the free-free luminosity at 2~cm (middle) and the synchrotron luminosity at 2~cm (right).

\section{Results}

Modelling the SED of NGC~891 in the wavelength range from 3.4~$\mu$m to 6.3~cm allows us to accurately determine the dust content, taking into account emission from the aromatic features as well as from the continuum emission from small and big grains, but also the radio emission composed by the free-free and the synchrotron radiation. 
Our model predicts a total dust mass of (4.18~$\pm$~0.21)~$\times~10^7$~M$_\odot$, while the mass fraction of the small carbon grains over the total dust mass is 0.10~$\pm$~0.01. The total radio luminosity at 2~cm is (5.54~$\pm$~0.24)~$\times~10^4$~L$_\odot$ with a fraction of 0.16~$\pm$~0.11 originating in free-free radiation. The inferred value of the synchrotron radiation index \textbf{($\mathbf{\alpha_s}$)} is 0.78~$\pm$~0.08 (in agreement with other studies, e.g. \cite{mulcahy}).

Having modelled the SED we can now study the largely unexplored wavelength range between $\sim 1$~mm and $\sim 2$~cm where a mix of emission mechanisms exists.
In the right panel of Fig.~\ref{fig2} we present the percentages to the global emission of dust (orange color), free-free (red color) and synchrotron emission (blue color) at 1, 2, 3, 5, 7~mm and 1, and 2~cm. From this plot it is evident that the dust thermal emission dominates to up to $\sim 3$~mm where 
the contributions of the free-free and synchrotron emission mechanisms become almost equally significant to the dust emission. At longer wavelengths, these latter emission mechanisms dominate with the synchrotron emission increasing rapidly with values up to $\sim 84\%$ at 2~cm, while the free-free emission reaches a maximum contribution of $\sim 26\%$ at $\sim 7$~mm  and then drops again. 

\begin{figure}[t]
\begin{center}
\includegraphics[scale=0.4]{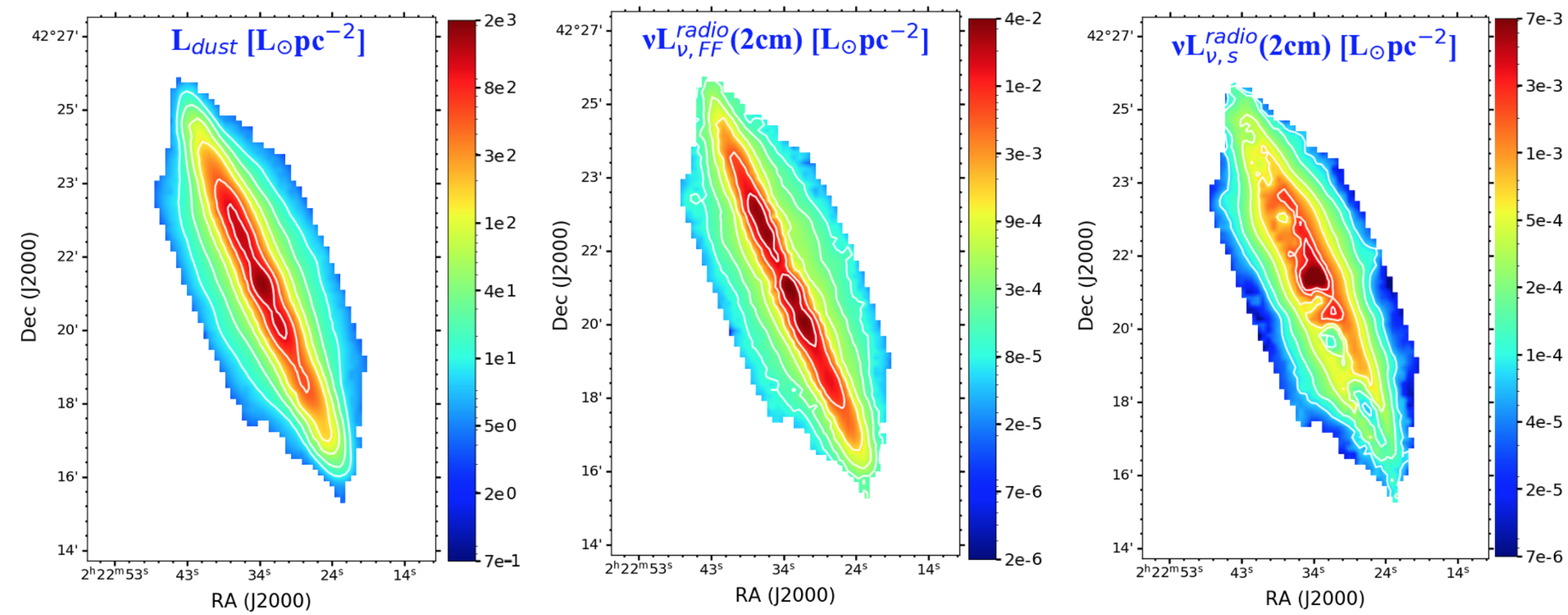}
\caption{Maps of the model-inferred luminosity for the dust emission (left panel), the free-free luminosity at 2~cm (middle panel) and the synchrotron luminosity at 2~cm (right panel).The colorbar on the right in each panel indicates the luminosity in units of L$_\odot$pc$^{-2}$.}
\label{fig3}
\end{center}
\end{figure}

The spatial decomposition of the emission of the galaxy (see Fig.~\ref{fig3}) allows us to determine the places where the different emission mechanisms dominate. From the dust luminosity map (left panel of Fig.~\ref{fig3}) it is evident that the bulk of the dust emission comes from the disc of the galaxy, as it is clearly seen also from the NIKA2 observations (Fig.~\ref{fig1}), but with significant amount also coming from the dust halo at large distances from the disc. The free-free emission at 2~cm (middle panel of Fig.~\ref{fig3}) shows a similar distribution to that of the dust emission, while the synchrotron luminosity (right panel of Fig.~\ref{fig3}) shows a different distribution with a deficiency along the galactic plane and with the peak emission originating from the galactic center. 
One interesting difference between the dust and the radio emission maps is the shape of the halo which, in the dust case has an elliptical shape, while in the radio case (and especially in the synchrotron component) has a more complex ``peanut''-like shape. 

\section{Conclusions}
\label{sec:concl}
In this study we present new millimetre NIKA2 observations of the edge-on galaxy NGC~891 at 1.15 and 2~mm.
Using the NIKA2 observations and also multiwavelength maps (from the mid-IR to the cm wavelengths) we construct the galaxy's SED and perform SED fitting analysis using the HerBIE SED fitting code. The observations indicate a morphology similar to that of the cold dust (detected at sub-mm) and of the molecular gas with the emission peaking at the nucleus of the galaxy and with two secondary maxima at $\sim 3$~kpc either side of the center. 
Emission at the NIKA2 bands is dominated by thermal dust, while the radio emission becomes almost equally significant to the thermal emission at $\sim 3$~mm. At longer wavelengths the radio emission dominates with synchrotron emission reaching up to $\sim 84\%$ at 2~cm. On local scales our analysis shows that the bulk of the dust thermal emission and the free-free emission comes from the disc of the galaxy while there is a deficiency, of the synchrotron emission, along the major axis of the disc of the galaxy.

\section*{Acknowledgements}
We would like to thank the IRAM staff for their support during the campaigns. The NIKA2 dilution cryostat has been designed and built at the Institut N\'eel. In particular, we acknowledge the crucial contribution of the Cryogenics Group, and in particular Gregory Garde, Henri Rodenas, Jean Paul Leggeri, Philippe Camus. This work has been partially funded by the Foundation Nanoscience Grenoble and the LabEx FOCUS ANR-11-LABX-0013. This work is supported by the French National Research Agency under the contracts "MKIDS", "NIKA" and ANR-15-CE31-0017 and in the framework of the "Investissements d’avenir" program (ANR-15-IDEX-02). This work has benefited from the support of the European Research Council Advanced Grant ORISTARS under the European Union's Seventh Framework Programme (Grant Agreement no. 291294). F.R. acknowledges financial supports provided by NASA through SAO Award Number SV2-82023 issued by the Chandra X-Ray Observatory Center, which is operated by the Smithsonian Astrophysical Observatory for and on behalf of NASA under contract NAS8-03060. This work was supported by the Programme National "Physique et Chimie du Milieu Interstellaire" (PCMI) of CNRS/INSU with INC/INP co-funded by CEA and CNES.

%
%
%

\end{document}